\documentclass[num-refs]{wiley-article}

\usepackage{siunitx}

\title{Graph Neural Network Predictions of Metal Organic Framework CO$_2$ Adsorption Properties}

\author[1,2\authfn{1}]{Kamal Choudhary}
\author[3]{Taner Yildirim}
\author[4]{Daniel W. Siderius}
\author[5]{A. Gilad Kusne}
\author[5]{Austin McDannald}
\author[5]{Diana L. Ortiz-Montalvo}

\affil[1]{Materials Science and Engineering Division, National Institute of Standards and Technology, Gaithersburg, 20899, MD, USA. }
\corremail{kamal.choudhary@nist.gov}
\affil[2]{Theiss Research, La Jolla, 92037, CA, USA. }
\affil[3]{NIST Center for Neutron Research, National Institute of Standards and Technology, Gaithersburg, 20899, MD, USA. }
\affil[4]{Chemical Sciences Division, National Institute of Standards and Technology, Gaithersburg, 20899, MD, USA. }

\affil[5]{Materials Measurement Science Division, National Institute of Standards and Technology, Gaithersburg, 20899, MD, USA. }

\corraddress{National Institute of Standards and Technology, Gaithersburg, 20899, MD, USA.}
\corremail{kamal.choudhary@nist.gov}
\fundinginfo{National Institute of Standards and Technology, Gaithersburg, 20899, MD, USA.}

\begin{document}

\begin{frontmatter}
\maketitle
\begin{abstract}
The increasing CO$_2$ level is a critical concern and suitable materials are needed to capture such gases from the environment. While experimental and conventional computational methods are useful in finding such materials, they are usually slow and there is a need to expedite such processes. We use Atomistic Line Graph Neural Network (ALIGNN) method to predict CO$_2$ adsorption in metal organic frameworks (MOF), which are known for their high functional tunability. We train ALIGNN models for hypothetical MOF (hMOF) database with 137953 MOFs with grand canonical Monte Carlo (GCMC) based CO$_2$ adsorption isotherms. We develop high accuracy and fast models for pre-screening applications. We apply the trained model on CoREMOF database and computationally rank them for experimental synthesis. In addition to the CO$_2$ adsorption isotherm, we also train models for electronic bandgaps, surface area, void fraction, lowest cavity diameter, and pore limiting diameter, and illustrate the strength and limitation of such graph neural network models. For a few candidate MOFs we carry out GCMC calculations to evaluate the deep-learning (DL) predictions.
\end{abstract}

\end{frontmatter}

\section{Introduction}
The CO$_2$ concentration in air is increasing at an alarming rate which could escalate global temperature by 1.5 °C by 2100 \cite{kirtman2013near}. CO$_2$ capture from pre-combustion plants is an effective way to reduce CO$_2$ level. Typical materials used in this process are zeolites, calcium chabazite, hydrotalcite, and activated carbon \cite{lee2011high}, but they usually suffer from low CO$_2$ working capacities, low tunability, and limited material diversity \cite{stocker2017characterization}.

Metal organic frameworks (MOFs) are a class of porous materials with inorganic clusters and organic building blocks, which can be repeated to build an extended 3D structure. MOFs provide high thermal stabilities (as high as 500 °C), a wide range of porosities (0.3–0.9), large surface areas (>8000 m$^2$g$^{-1}$), low densities (0.2 g cm$^{-3}$), a wide range of pore sizes (3--100\,Å), active sites that can be used for site-specific adsorption and catalysis, and most importantly a wide range of chemical tunability unlike conventional porous materials \cite{altintas2021machine}. Hence, MOFs are considered as one of the top ten materials for emerging technologies by the International Union of Pure and Applied Chemistry (IUPAC) \cite{gomollon2019ten}.

In the past, there have been several efforts to use conventional computational methods such as grand canonical Monte Carlo (GCMC) to develop CO$_2$ isotherm databases for thousands of hypothetical as well as experimentally realizable MOFs \cite{altintas2018database,chung2019advances,bobbitt2016high,wilmer2012large,wilmer2012structure,sikora2012thermodynamic,colon2017topologically,gomez2016evaluating,rosen2021machine}. However, such computational methods are still time consuming, especially to screen thousands to millions of MOF candidates. One of the most obvious ways to tackle this issue is using machine learning (ML) methods, where ML methods act as a pre-screening tool for conventional computational methods, which then can be used as a screening tool for experimental designs \cite{vasudevan2019materials,schleder2019dft,schmidt2019recent,ohno2016machine,fernandez2014rapid,jablonka2020big,chong2020applications,jablonka2020big}.

The key precursors for applying ML methods for CO$_2$ absorption in MOF are : 1) large datasets of experimental or computational CO$_2$ isotherms and other MOF properties, 2) an ML framework that can accurately represent arbitrary MOF crystals and chemical structures. Fortunately, due to materials genome initiative (MGI) \cite{de2019new} and similar initiatives there has been a rapid development for generating large databases of MOFs such as: Computation-Ready, Experimental (CoREMOF), hypothetical (hMOF), and quantum (QMOF) \cite{chung2019advances,bobbitt2016high,wilmer2012large,wilmer2012structure,sikora2012thermodynamic,colon2017topologically,gomez2016evaluating,rosen2021machine}. These databases are based on computational methods such as density functional theory (DFT) or GCMC and the results are compared to experiments wherever applicable \cite{sun2014computational}. They contain exact coordinates for MOF atomic structures and their properties. 

Machine learning has been already used for some of the atomistic property predictions of MOFs \cite{altintas2021machine,anderson2018role,guda2020machine,burner2020high,fernandez2014rapid,chong2020applications,yao2021inverse}, but the application of multi-output ML methods especially for pressure dependent CO$_2$ adsorption has not been reported earlier to the best of our knowledge. Also, previous ML models are based on cumbersome handcrafted features which can be superseded with deep learning based methods \cite{kearnes2016molecular}. Graph neural network (GNN)/deep learning (DL) methods such as Crystal Graph Convolutional Neural Networks (CGCNN) \cite{xie2018crystal} and Atomistic Line Graph Neural Network (ALIGNN) \cite{choudhary2021atomistic} can be used to represent arbitrary chemistry and atomic structure with respect to interatomic bonds and angles and can utilize the full power of deep neural network framework.

In this work, we apply the ALIGNN model for predicting properties in hMOF and QMOF databases which can be used for accelerated MOF design. We train models for technologically important quantities such as CO$_2$ adsorption isotherms at different pressures, electronic bandgap, lowest cavity diameter (LCD), pore limiting diameter (PLD), void fraction, volumetric, and gravimetric surface areas. After training the models, we apply them to more CoREMOF database to screen for potential  MOF candidates for CO$_2$ capture. We compare the model predictions with a few experimental measurement results to check accuracy. Additionally, we identify some of the key strength and challenges of the DL framework.

\section{Methods}

This study uses the QMOF dataset \cite{rosen2021machine} and the hMOF database \cite{bobbitt2016high} to train the DL model. The trained DL model is applied to the CoreMOF database \cite{chung2019advances} to computationally rank the MOFs based on max CO$_2$ adsorption. 

The QMOF dataset consists of 20425 MOFs that Rosen \textit{et al.} \cite{rosen2021machine} generated using a subset of the CoRE-MOF and Cambridge Structural Database (CSD) \cite{moghadam2017development} and made them suitable for density functional theory (DFT) calculations.  The DFT calculations \cite{nistdisclaimer} were carried out at PBE-D3(BJ) level with Vienna \textit{ab initio} Simulation Package (VASP) \cite{kresse1996efficient,kresse1999ultrasoft}.

The hMOF database consists of 137953 MOF structures which were generated using 102 building blocks from Wilmer \textit{et al.} \cite{wilmer2012large}. Details about the generation of the hMOF database can be found in Wilmer \textit{et al.} \cite{wilmer2012structure}. Briefly, the Grand canonical Monte-Carlo (GCMC) calculations were performed using the RASPA code \cite{dubbeldam2016raspa}. GCMC requires reliable force-fields for simulation. Interaction energies between non-bonded atoms were computed through the Lennard-Jones (LJ) and Coulomb potentials. LJ parameters for the framework atoms were taken from the Universal Force Field (UFF) \cite{rappe1992uff}. Partial charges and LJ parameters for CO$_2$ were taken from the TraPPE force field \cite{potoff2001vapor}. Each GCMC calculation used a 1000 cycle equilibration period followed by a 1000 cycle production run. A cycle consists of \textit{n} Monte Carlo steps; where \textit{n} is equal to the number of molecules (which fluctuates during a GCMC simulation). All
simulations included random insertion, deletion, rotation, and translation moves of molecules with equal probabilities.

Unlike the previous database where the MOF structures have been generated using computer algorithms, Computation-Ready, Experimental Metal–Organic Framework (CoREMOF) database contains materials that are derived from synthesized materials following the synthesis and data-refining protocols to provide closer to reality structures. CoREMOF database contains more than 14000 MOF structures \cite{chung2019advances}. There is no direct overlap between hMOF and CoREMOF database.

Graph neural networks are trained with atomistic line graph neural network (ALIGNN) \cite{choudhary2021atomistic} which is available at \url{https://github.com/usnistgov/alignn}. ALIGNN has been used to train more than 50 properties of solids and molecules with high accuracy. A MOF structure is represented as a graph using atomic elements as nodes, and atomic bonds as edges. Each node in the atomistic graph is assigned 9 input node features based on its atomic species:
electronegativity, group number, covalent radius, valence electrons, first ionization energy, electron affinity, block and atomic volume. The inter-atomic bond distances are used as edge features with radial basis function upto 8 $\textrm{\AA}$ cut-off. We use a periodic 12-nearest-neighbor graph construction. This atomistic graph is then used for constructing corresponding line graph using interatomic bond-distances as nodes and bond-angles as edge features. ALIGNN uses Edge-gated graph convolution for updating nodes as well as edge features. One ALIGNN layer composes an edge-gated graph convolution on the bond graph with an edge-gated graph convolution on the line graph. The line graph convolution produces bond messages that are propagated to the atomistic graph, which further updates the bond features in combination with atom features. The ALIGNN model is implemented in PyTorch \cite{paszke2019pytorch} and deep graph library (DGL) \cite{wang2019deep}.

For regression targets we minimize the mean squared error (MSE) loss for 50 epochs using the AdamW optimizer with normalized weight decay of 10$^{-5}$ and a batch size of 64.The learning rate is scheduled according to the one-cycle policy with a maximum learning rate of 0.001. We use 80 initial bond radial basis function (RBF) features, and 40 initial bond angle RBF features. The atom, bond, and bond angle feature embedding layers produce 64-dimensional inputs to the graph convolution layers. The main body of the network consists of 4 ALIGNN layers and 4 graph convolution (GCN) layers, each with hidden dimension 256 nodes. The final atom representations are reduced by atom-wise average pooling and mapped to regression or classification outputs by a single linear layer. 

We used National Institute of Standards and Technology (NIST)’s Nisaba cluster to train all ALIGNN models \cite{nistdisclaimer}. Each model is trained on a single Tesla V100 SXM2 32 gigabyte Graphics processing unit (GPU), with 8 Intel Xeon E5-2698 v4 CPU cores for concurrently fetching and preprocessing batches of data during training. We use 80 \%:10 \%: 10 \% splits. The 10 \% test data is never used during training procedures. Except for the pressure dependent CO$_2$ adsorption data, which is a multi-output model, all other models are single output regression.

\section{Results and discussion}
We apply ALIGNN model to predict MOF properties in QMOF and hMOF databases. QMOF is developed using quantum density functional theory method while hMOF is based on classical force field guided GCMC calculations. The QMOF and hMOF materials are composed of some of the combination of the following 9 elements: C, H, O, N, Zn, Cu, Zr, F, Cl, and Br. These datasets are kept in the jarvis-tools package  \url{https://jarvis-tools.readthedocs.io/en/master/databases.html} \cite{choudhary2020joint} in a format to easily apply DL methods. In this section, we will first visualize the data distribution and then discuss the ALIGNN application on these datasets.

\subsection{Application to QMOF database}
\begin{figure}[hbt!]
    \centering
    \includegraphics[trim={0 0.1cm 0 .1cm},clip,width=1.0\textwidth]{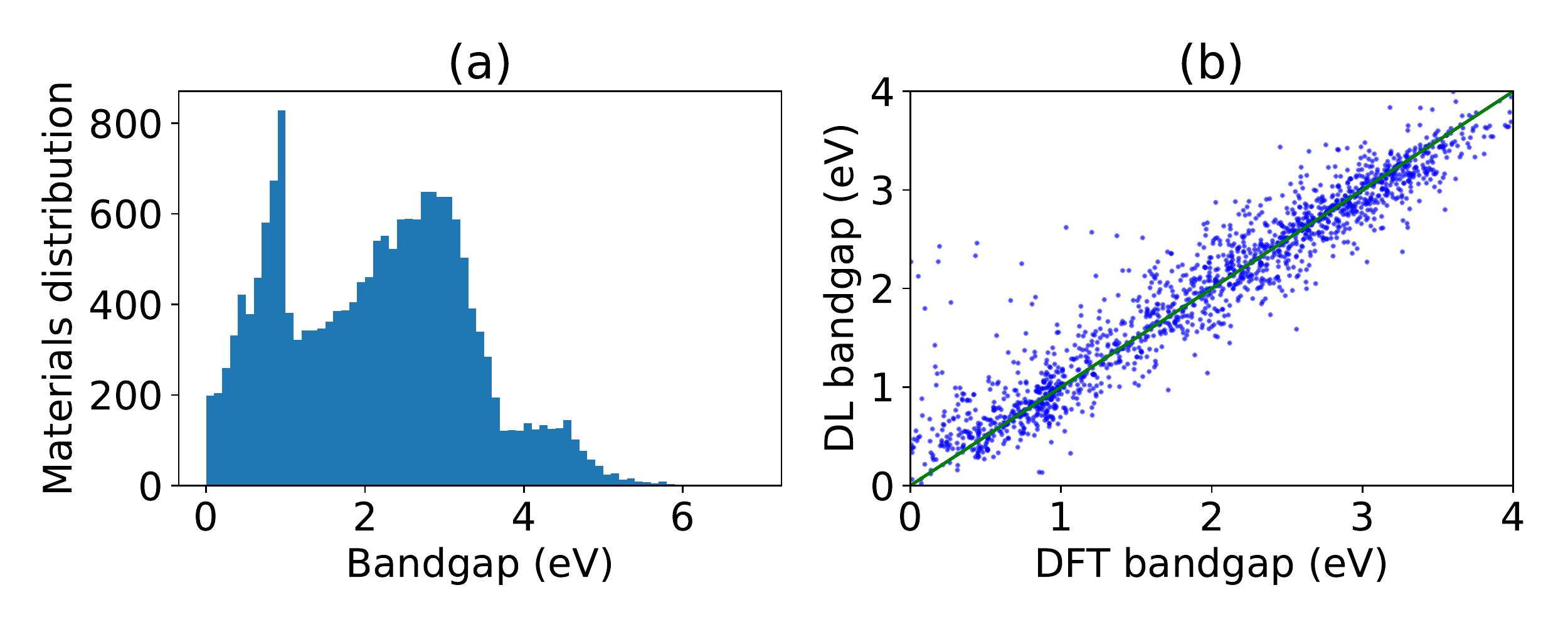}
    \caption{Data visualization of QMOF electronic bandgaps and GNN predictions on the test set. a) Electronic bandgap data distribution in QMOF bandgap distribution database, b) ALIGNN based deep-learning (DL) predictions on 10 \% test set.}
\end{figure}

Before applying the DL model on QMOF we visualize the properties in QMOF database with about 20425 MOFs. One of the unique features of QMOF is electronic bandgap data which cannot be obtained from classical force field calculations such as in hMOF.  Although the focus of the work is on CO$_2$ adsorption, we note that MOFs can be viewed as photocatalysts and tailoring of bandgaps can provide dual technological benefits of CO$_2$ reduction as well as solar power generation \cite{lee2014cu,guo2021band}. We visualize the bandgap data distribution in Fig. 1a. We observe a high peak around 1 eV and 3 eV with data range up to 6 eV.

The developers of QMOF applied several machine-learning models for predicting the bandgaps. They compared well-known machine learning model such as Sine Coulomb matrix, Orbital field matrix \cite{rupp2012fast,bartok2013representing}, Smooth overlap of atomic orbitals (SOAP) \cite{de2016comparing}, and crystal graph convolution neural network (CGCNN) \cite{xie2018crystal} to develop a model for predicting the bandgap of MOFs. They found that the best model was achieved with CGCNN with a mean absolute error (MAE) of 0.274 eV for bandgap prediction. Following a similar training strategy, we trained ALIGNN model on this dataset and find the MAE to be 0.20 eV, which is nearly 27 \% improvement in accuracy. A parity plot for the actual DFT data and ALIGNN based predictions on 10 \% test set is shown in Fig. 1b. For a perfect agreement, all the data-points would fall upon $y = x$ line, which the data points in Fig. 1b closely resemble. ALIGNN has already been shown to outperform CGCNN and other models for other solid-state and molecular dataset and the previous results support the findings \cite{choudhary2021atomistic}.

\subsection{Application to hMOF database}

\begin{figure}[hbt!]
    \centering
    \includegraphics[trim={0 .1cm 0 .1cm},clip,width=1.0\textwidth]{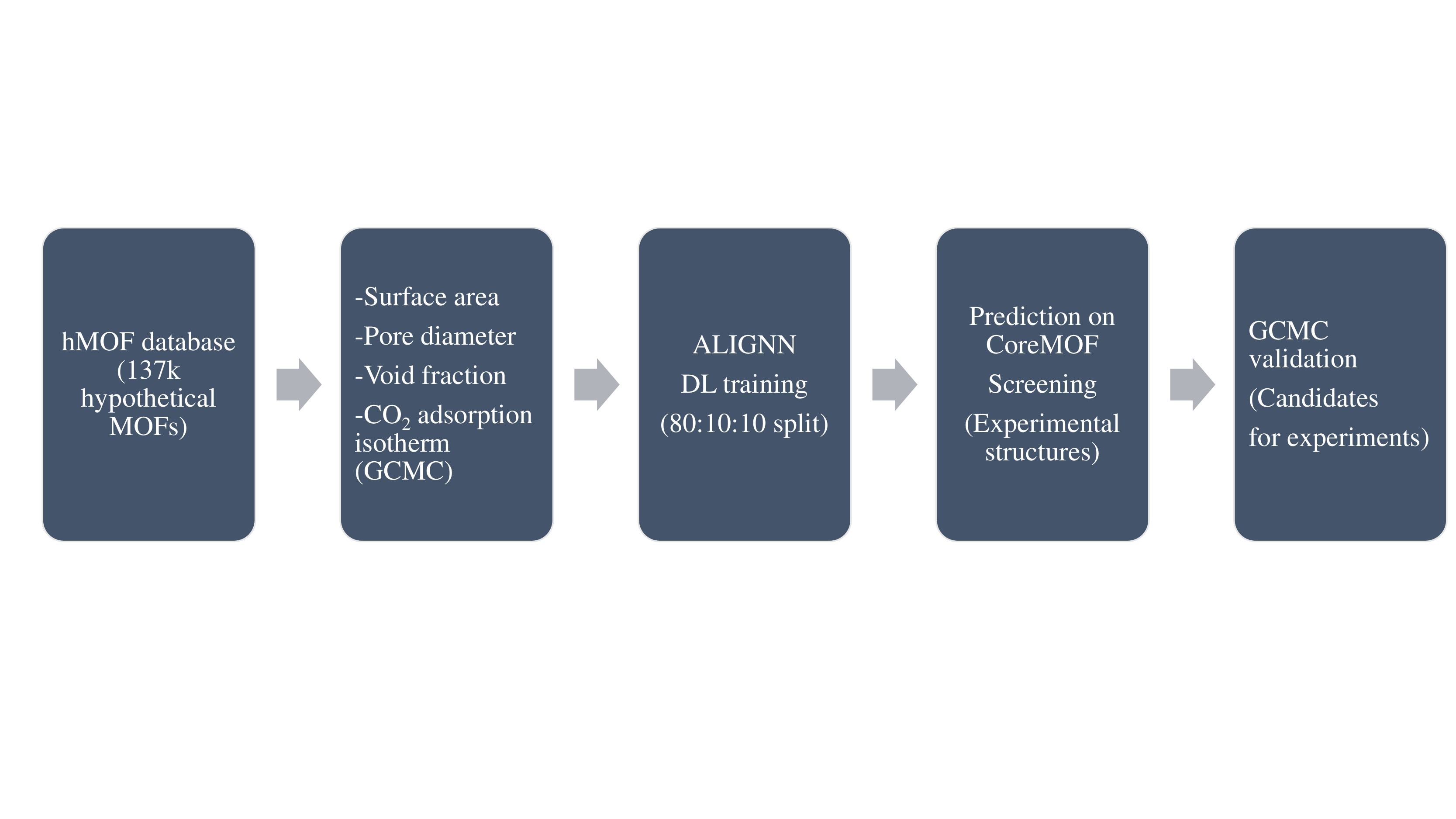}
    \caption{Workflow used in training and applying graph neural network models for several properties in hMOF database.}
\end{figure}
Next we apply the ALIGNN model for training properties in the hMOF databases. The list of properties include: gravimetric surface area ( m$^2$g$^{-1}$) , volumetric surface area (m$^2$cm$^{-3}$), largest cavity diameter (LCD), pore limiting diameter (PLD), void fraction, and pressure dependent CO$_2$ absorption. All the models except the CO$_2$ isotherm are single output models. The adsorption isotherm predictions has 5 outputs representing CO$_2$ absorption at 5 different pressures 0.01, 0.05, 0.1, 0.5, and 2.5 bar. In addition to the multi-output model, we also train models for minimum (0.01 bar) and maximum (2.5 bar) pressures (under consideration) to check whether having more data as targets (five versus one) helps improve the model predictions.
\begin{figure}[hbt!]
    \centering
    \includegraphics[trim={0 .1cm 0 .1cm},clip,width=1.0\textwidth]{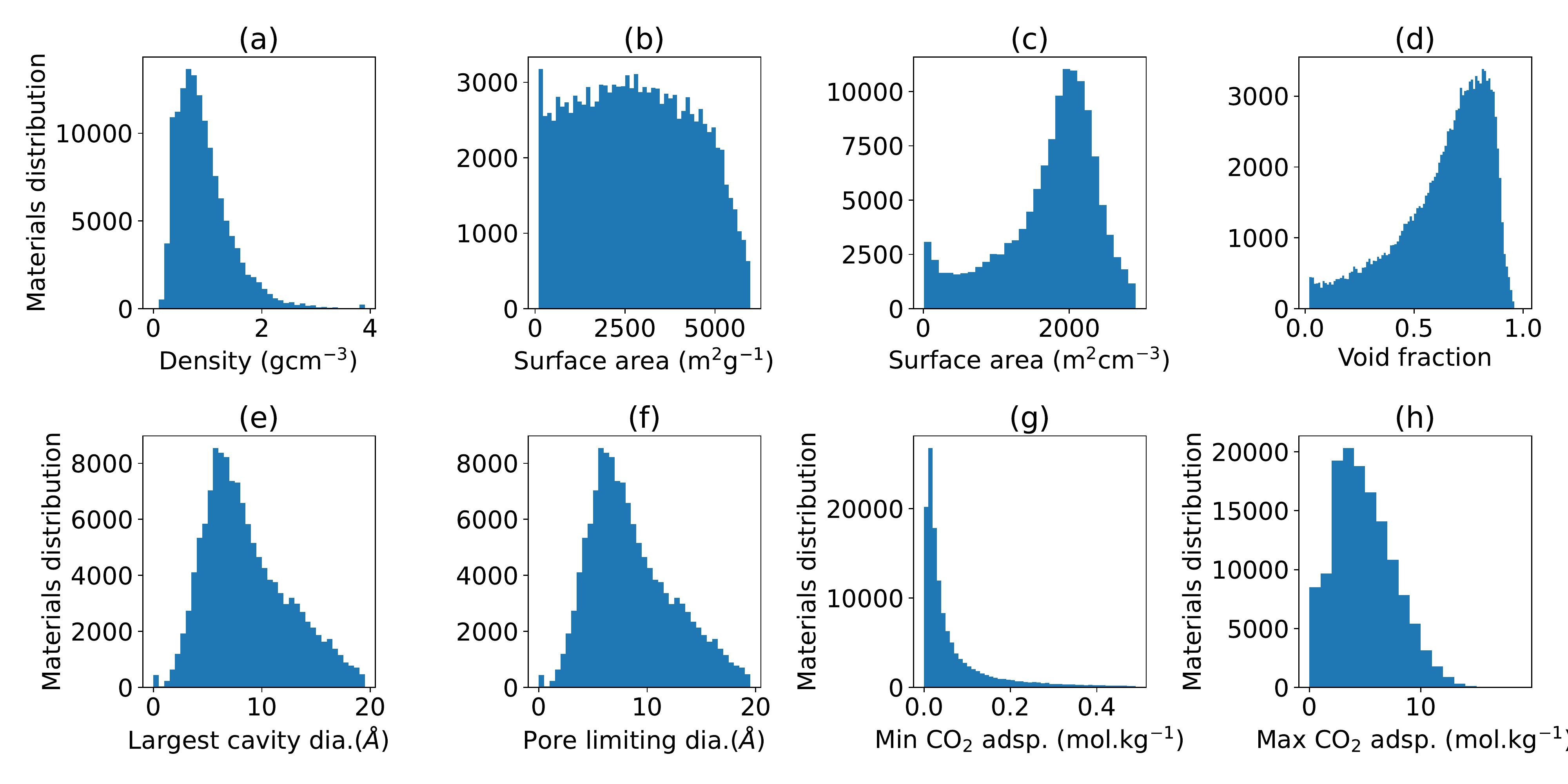}
    \caption{Data distribution visualization of several properties in the hMOF database. a) density, b) gravimetric surface area, c) volumetric surface area, d) void fraction, e) LCD, f) PLD, e) minimum value of CO$_2$ dsorption among 5 pressure values (i.e. adsorption at 0.0 1bar), and f) maximum value of CO$_2$ adsorption (i.e. adsorption at 2.5 bar) among 5 pressure values.}
\end{figure}
\begin{figure}[hbt!]
    \centering
    \includegraphics[trim={0 .1cm 0 .1cm},clip,width=1.0\textwidth]{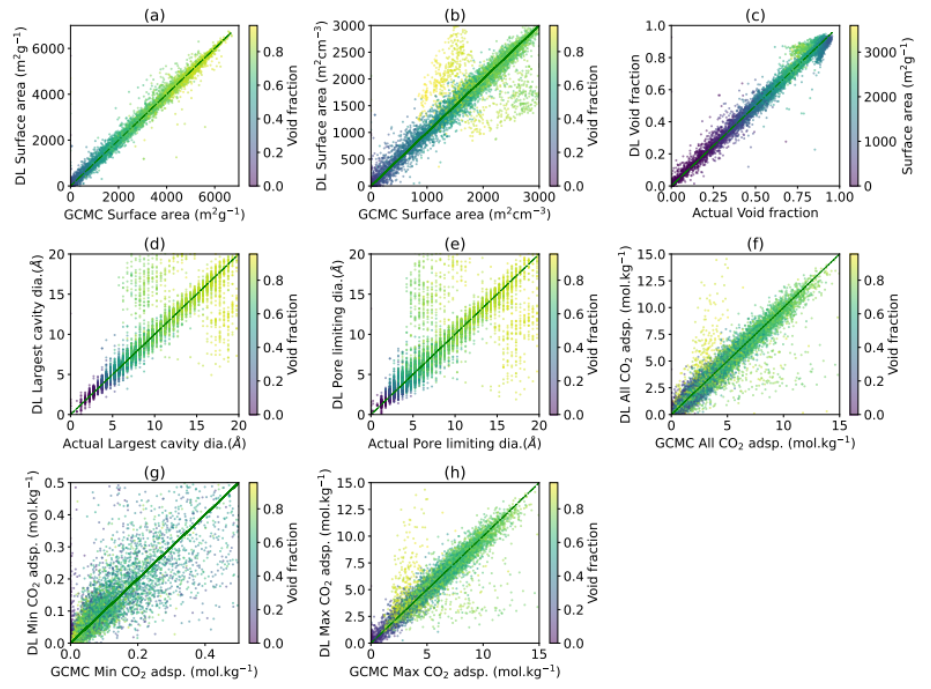}
    \caption{GNN predictions and GCMC actual value comparisons of several properties on 10 \% held test set (13796 MOFs) in the hMOF dataset. a) gravimetric surface area, b) volumetric surface area, c) void fraction, d) largest cavity diameter, e) pore limiting diameter, f) adsorption at five different pressures, g) adsorption at 0.01 bar, and h) adsorption at 2.5 bar, }
\end{figure}

In Fig. 2 we show the training process for hMOF which is described below. 
Before applying any DL, we first visualize the distribution of the above mentioned properties for 137953 MOFs in Fig. 3. As observed in Fig. 3a, MOFs in hMOF have low bulk densities with a peak around 0.7 g cm$^{-3}$, which is much lower than usual solids such as silicon (2.3 g cm$^{-3}$) and aluminum (2.7 g cm$^{-3}$). The gravimetric and volumetric surface areas in Fig. 3b and Fig. 3c varies up to 6000 m$^2$ g$^{-1}$ and 3000 m$^2$ cm$^{-3}$, respectively. The gravimetric surface area is almost uniformly distributed across the range while the volumetric surface area peaks around 2000 m$^2$ cm$^{-3}$. We note that high volumetric surface area doesn't necessarily imply high gravimetric surface area and vice versa. The void fraction in Fig. 3d shows a peak around 0.8 suggesting high porosity of MOFs. The largest cavity diameters (LCD) and pore limiting diameters (PLD) are other critical parameters for adsorption. Both of them vary up to 20 $\textrm{\AA}$ with a peak around 7 $\textrm{\AA}$ (as shown in Fig. 3e and Fig. 3f). As mentioned above, hMOF contains CO$_2$ adsorption at 5 different and monotonically increasing pressure values, so we show the CO$_2$ adsorption at 0.01 bar and 2.5 bar pressures in Fig. 3g and Fig. 3h, respectively.

We note that some of these quantities (\textit{e.g.} PLD, LCD, and void fraction) can be easily calculated for the whole dataset of potential MOFs using conventional computational methods such as Zeo++ \cite{willems2012algorithms}. Using DL based methods to predict these properties will elucidate the strengths and weaknesses of the DL models. However, predicting adsorption isotherms can be prohibitively expensive for the whole dataset by traditional means (\textit{e.g.} GCMC). Therefore the DL methods can be useful as a surrogate model, with the usefulness characterized by the appropriate performance metrics. 

The parity plots for predictions on the test dataset by the ALIGNN models for different properties is shown in Fig 4. For an ideal model the DL and actual target values should line on $y = x$ straight line. Corresponding performance metric for the trained models are charted in Table. 1. In addition to the performance metric such as the mean absolute error (MAE), we also provide the mean absolute deviation (MAD) which serves as the baseline for the model predictions. Usually, a good machine learning model has MAE five times lower than its MAD. The MAD:MAE ratio also serves as a good comparison metric across different properties. From Table 1, we observe that most of our models have MAD:MAE greater than 5, with highest value for gravimetric surface area and the minimum value for the pore limiting diameter. Same behavior can be observed from the parity plots. 

Comparing Fig. 4a and Fig. 4b, we observe that gravimetric surface area is easier to predict than the volumetric surface area. Currently we don't have a good explanation for this behavior. The void fraction is easier to predict as well as shown in Fig. 4c. The LCD and PLD parity plots in Fig, 4d and Fig. 4e suggest they are much difficult to predict. This shows one of the limitations of the ALIGNN model where it cannot capture long range order based properties. In Fig. 4f, we show the DL predicted data against the actual GCMC data for all the pressures in one plot. Similar plots for the minimum and maximum pressures are shown in Fig. 4 g and 4h. Interestingly, the full-isotherm data based model has lower overall MAE as well as lower MAD:MAE in comparison with the maximum pressure adsorption model. This result suggests that training multi-output model could be advantageous for adsorption data especially at higher pressure values. The colorbar in the plots represent void fraction with the exception of Fig. 4c (surface area is represented). We observe that most of the deviations from $y = x$ line in the test set occur on high void fraction MOFs (such as upper yellow part of Fig. 4b, 4d, 4e and 4h). This can be due to the k-nearest neighbor and relatively shorter cut-off (8 $\textrm{\AA}$) strategies for graph construction in ALIGNN models. Larger k-nearest neighbors and cut-off impose higher computational expense, a challenge for ALIGNN. 

In addition to the parity plots, we also analyze the best and worst isotherm predictions using the multi-output isotherm predictor model. We sort the test data based predictions with respect to MAE and show top 5 results in Fig. 5a to Fig. 5e and worst values from Fig. 5f to Fig. 5j. Clearly, the best predictions are on par with the target data. For the worst predictions, the multi-output predictions are higher in some cases and lower in others compared to actual GCMC data implying non-uniform trends.

\begin{figure}[hbt!]
    \centering
    \includegraphics[trim={0 .1cm 0 .1cm},clip,width=1.0\textwidth]{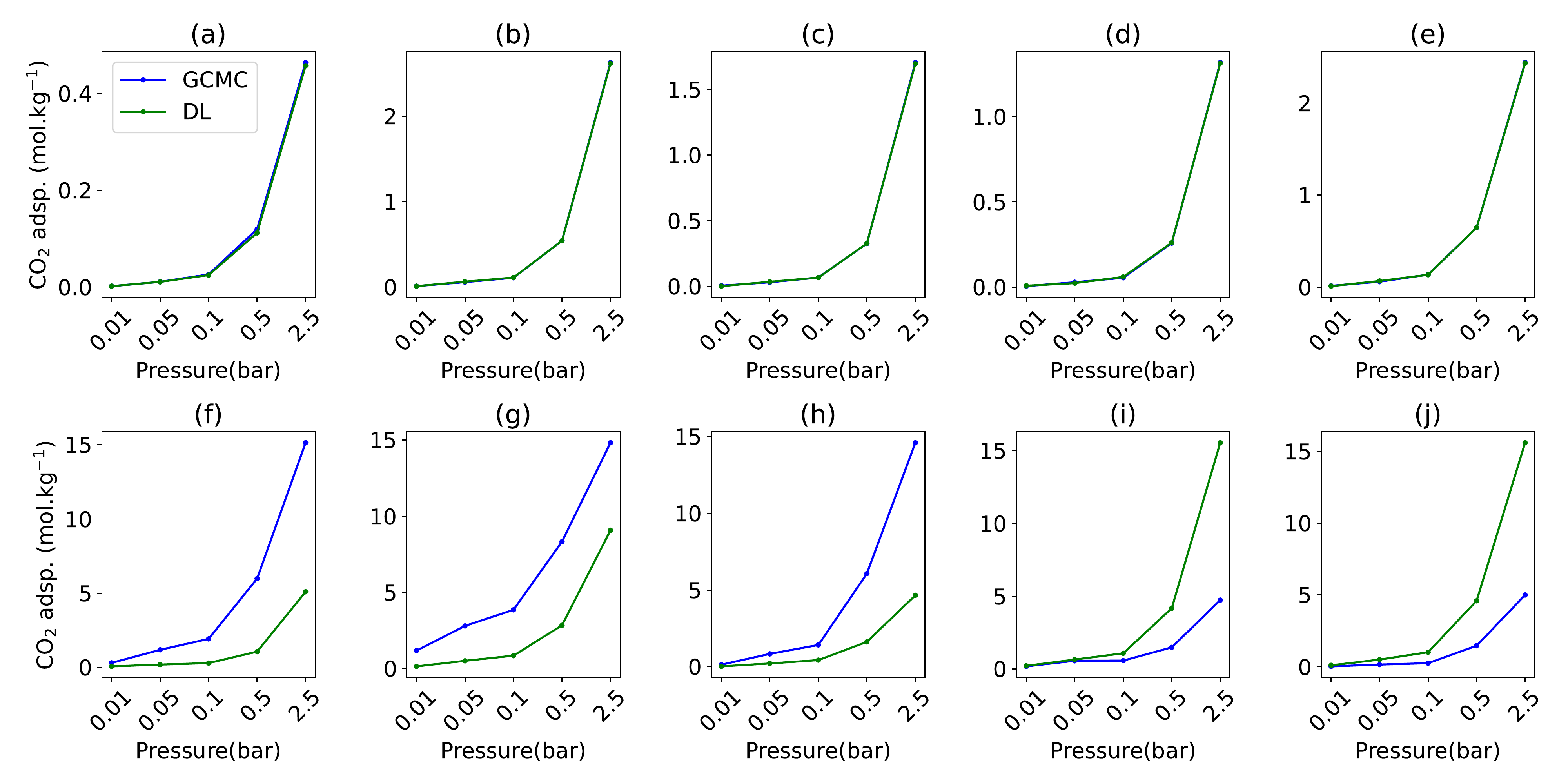}
    \caption{Comparison of actual GCMC and ALIGNN predictions of CO$_2$ adsorption  at different pressures values. The model was tested on the 10 \% held set of 13796 materials, out of which 5 best and 5 worst predictions are shown. a-e) 5 best predictions, f-j) 5 worst predictions.}
\end{figure}

\begin{table}
\centering
\caption{ALIGNN performance on hMOF properties interms of mean absolute error (MAE) amd mean absolute deviation (MAD).}
\begin{minipage}{174pt}

\begin{tabular}{|c|c|c|c|c|}
Property & Unit & MAD  & MAE & MAD:MAE\\
Grav. surface area & m$^2$ g$^{-1}$ & 1430.82 & 96.06 & 14.90 \\
Vol. surface area & m$^2$ cm$^{-3}$ & 561.44 & 109.84 & 5.11\\
Void fraction & No unit & 0.16 & 0.018 & 8.89\\
LCD & $\textrm{\AA}$ & 3.44 & 0.76 & 4.53\\
PLD & $\textrm{\AA}$ & 3.55 & 0.92 & 3.86\\
All adsp & mol kg$^{-1}$ & 1.70 & 0.19 & 8.95\\
Min adsp & mol kg$^{-1}$ & 0.12 & 0.04 & 3.0\\
Max adsp & mol kg$^{-1}$ & 2.16 & 0.48 & 4.50\\
\end{tabular}
\end{minipage}
\end{table}

To investigate the reason why the DL model works well for a certain class of materials, we carried out several correlation analyses between the errors of multi-output isotherm predictor model and materials properties. We calculate the mean of absolute errors between actual and DL predictor values. From Table 2, we find the minimum and maximum CO$_2$ intake and density of the MOFs relate the most to the errors in the DL predictions. Note that there are relatively few examples of low-density materials, hence DL is less likely to generalize well for these set of materials.

\begin{table}[hbt!]
\centering
\caption{Pearson and Spearman correlation between mean of mean absolute errors of the CO$_2$ absorptions on test set and physical properties.}
\begin{minipage}{174pt}

\begin{tabular}{|c|c|c|c|c|}
Property & Unit & Pearson  & Spearman \\
Density & g m$^{-3}$ & 0.04 & 0.22 \\
Grav. surface area & m$^2$ g$^{-1}$ & -0.-07 & -0.10  \\
Vol. surface area & m$^2$ cm$^{-3}$ & -0.-04 & -0.10 \\
Void fraction & No unit & -0.02 & -0.2 \\
N atoms & No unit  &0.1 & 0.2 \\
PLD & $\textrm{\AA}$ & -0.07 & -0.23\\
LCD & $\textrm{\AA}$ & -0.05 & -0.21\\
Min adsp & mol kg$^{-1}$ & 0.33 & 0.56\\
Max adsp & mol kg$^{-1}$ & 0.27 & 0.40\\
\end{tabular}
\end{minipage}
\end{table}

Next, we evaluate the accuracy of the DL model based predictions by comparing them with a few experimental and GCMC calculation results. Over the years we have experimentally characterized a large number of MOFs using same measurement protocol and very accurate high-pressure sievert apparatus\cite{peng2013jacs}. 
We compare the experimental CO$_2$ adsorption at 2.5 bar for some of these previously well characterized MOFs \cite{simmon2011enenv,peng2013jacs,wang2021microporous,bucior2019energy} with DL predictions in Fig. 6. We observe excellent comparison between DL and experimental results. We find that the mean absolute difference for DL and experimental values for CO$_2$ adsorption and surface area are 1.54 molkg$^{-1}$ and 399.75 m$^2$g$^{-1}$ which are approximately 3 and 4 times higher than MAE of respective models in Table. 1. The above results suggest that the DL based model predictions can be useful for guiding experimental measurements.

Moreover,  we apply the DL model for predicting maximum CO$_2$ adsorption trained on hMOF database to screen high CO$_2$ adsobing MOFs in the CoREMOF database Note that there is almost no overlap between the MOFs in these two databases. The hMOF database is based on hypothetical structures while CoreMOF uses experimental structures which have been processed following several protocols. DL prediction for a MOF takes less than a few seconds. We rank the MOFs and provide the DL based predictions for MOFs in the supplementary information. Some of the high rank materials predicted by our DL model are: ATEYUV (10.7 mol.kg$^{-1}$), WEHJUQ (10.2 mol.kg$^{-1}$), ATEYOP (9.94 mol.kg$^{-1}$), HOJLID (9.86 mol.kg$^{-1}$), and MINCUJ (9.81 mol.kg$^{-1}$) , shown in Fig. 7 and Fig. 8. The crystal structure visualization for these MOFs as well as other screened materials are available at https://mof.tech.northwestern.edu

\begin{figure}[hbt!]
    \centering
    \includegraphics[trim={0 0.1cm 0 .1cm},clip,width=1.0\textwidth]{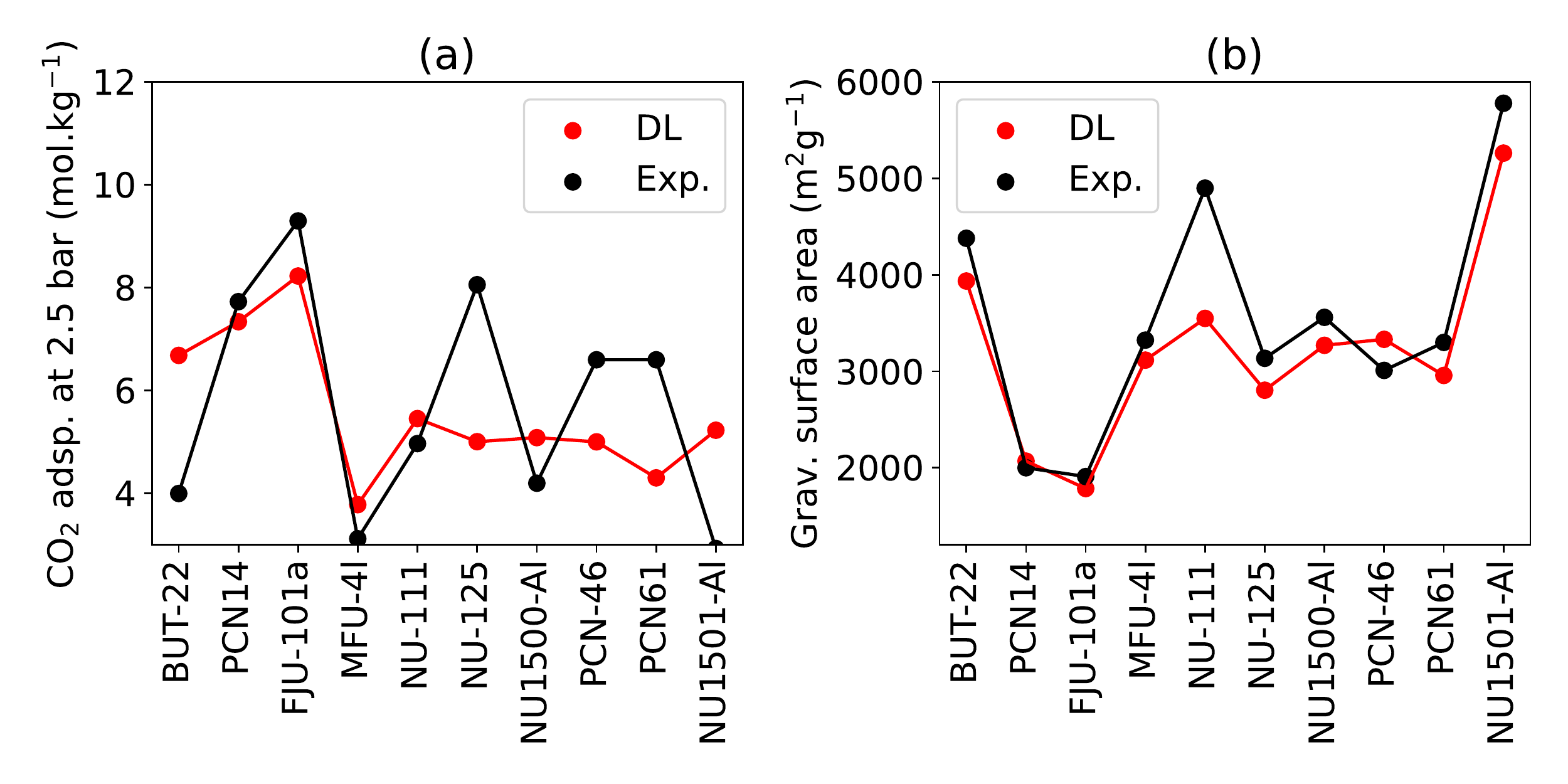}
    \caption{Comparison of experimental measurements and ALIGNN (DL) predictions for a few previously characterized MOFs from literature. a) CO$_2$ adsorption , b) gravimetric surface area.}
\end{figure}

\begin{figure}[hbt!]
    \centering
    \includegraphics[trim={0 .1cm 0 .1cm},clip,width=1.0\textwidth]{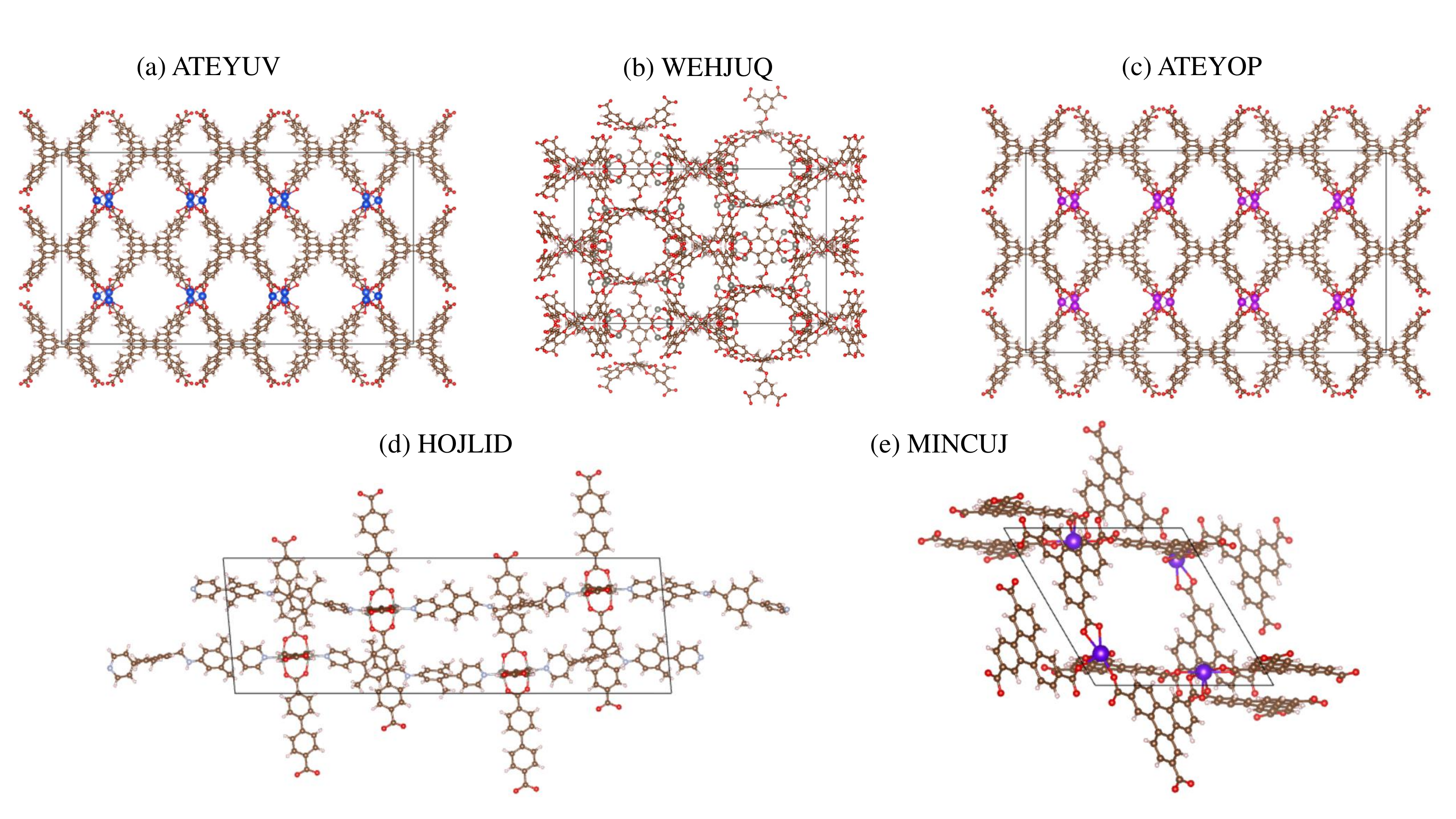}
    \caption{ALIGNN predicted some examples of the high-CO$_2$ adsorbing MOFs from CoREMOF DB. Complete list is provided in the supplementary information. a) ATEYUV, b) WEHJUQ, c) ATEYOP , d) HOJLID , and 
    e) MINCUJ. The crystal structure visualization for these MOFs as well as other screened materials are available at https://mof.tech.northwestern.edu.}
\end{figure}

\begin{figure}[hbt!]
    \centering
    \includegraphics[trim={0 0.1cm 0 .1cm},clip,width=1.0\textwidth]{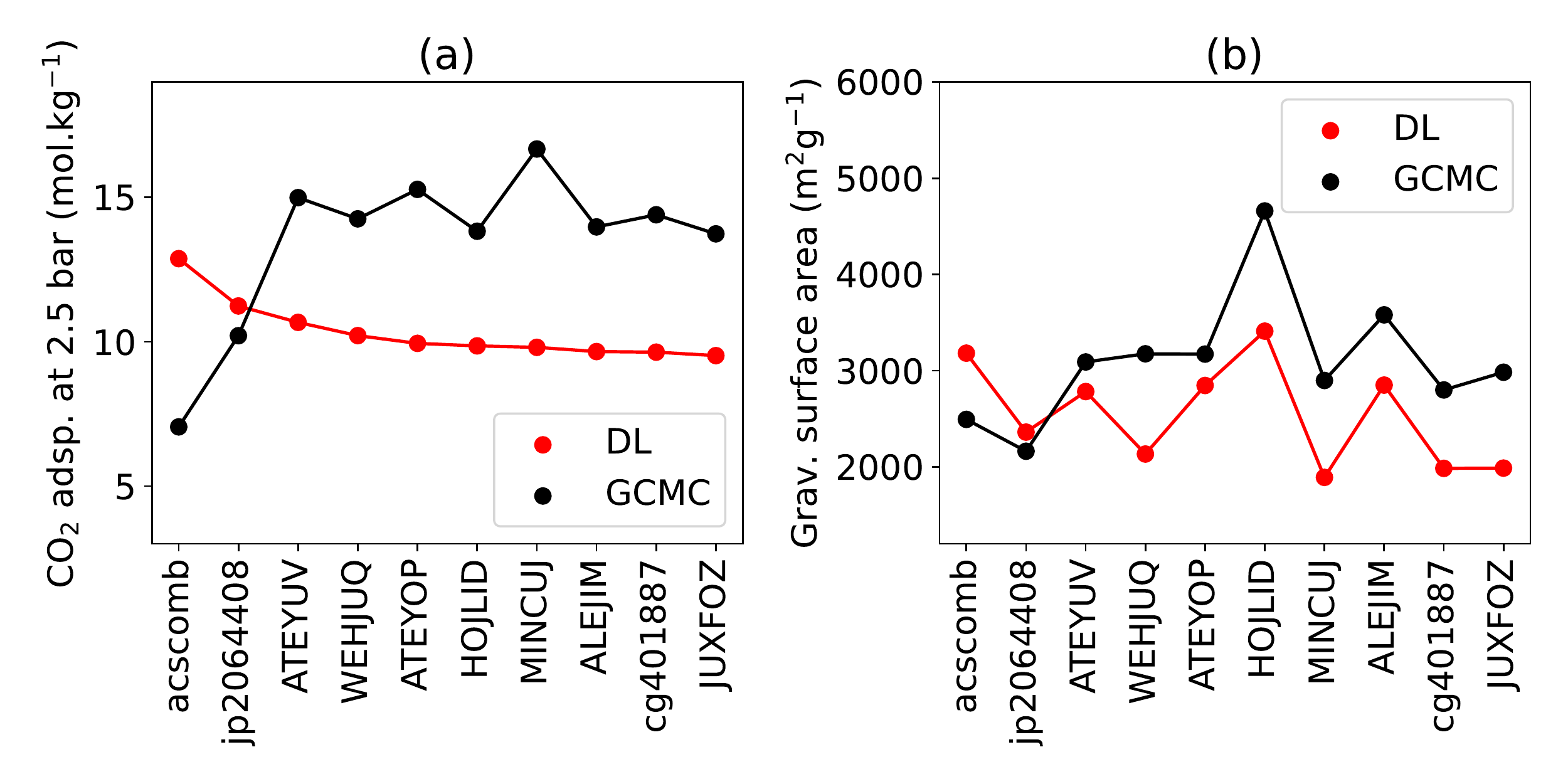}
    \caption{Comparison of GCMC and ALIGNN (DL) predictions for top 10 screened candidates from CoREMOF DB. a) CO$_2$ adsorption , b) gravimetric surface area.}
\end{figure}

In order to check the DL predictions, we carried out accurate GCMC simulations and compare the results with DL predictions. In Fig. 8a, the comparison of GCMC vs DL predictions for CO$_2$ on top 10 screened candidates shows the DL method can indeed predict high CO$_2$ adsorption MOFs. The GCMC data are obtained with the similar computation set-up as of hMOF database but with larger cutoff (up to 18 \AA) and
atomic charges of MOFs. The partial atomic charges of the framework  are obtained by message passing neural networks model developed in Ref.\cite{raza2020physchemC} and included in the CO$_2$ adsorption simulations. Although the GCMC values do not perfectly align with the DL data, the trends are noteworthy. We note that nine out of ten predicted structures give a CO$_2$ adsorption above 10 $mol/kg$ at 2.5 bar, which is quite high compared to all other experimentally well known MOFs. Fig. 8a clearly suggests that DL can predict high performing MOFs which are also validated with the GCMC calculations. We find that the mean absolute difference between the GCMC and DL values for CO$_2$ adsorption at 2.5 bar for these candidates is 4.2 molkg$^{-1}$ which is almost 9 times higher than that of MAE reported in Table 1. Next we compare the DL and GCMC bases surface area predictions for these top 10 screened MOFs. We notice that the trends in the predictions (as shown in Fig. 8b) are much better than that of CO$_2$ adsorption. We find the mean absolute difference between these values as 697.2 m$^2$g$^{-1}$ which is 7 times higher than MAE of the DL model obtained for gravimetric surface area in Table. 1. Experimental synthesis and characterizations of these is beyond the scope of the present work and will be carried out in future. The complete list of MOFs during the above screening procedure is provided in the supplementary information. The trained ALIGNN (DL) models for the properties trained above are also distributed in the ALIGNN GitHub repo as pre-trained models for making quick predictions on arbitrary MOFs.





\section{Conclusion}
In this article, we have applied atomistic line graph neural network (ALIGNN) to predict several properties such as volumetric and gravimetric surface area, void fraction, maximum and minimum CO$_2$ adsorption, adsorption isotherms at several pressures and electronic bandgaps in hMOF and QMOF datasets. High accuracy models are obtained for surface area and CO$_2$ adsorption that can be used for pre-screening applications. We have applied the trained model to pre-screening high-performance MOFs in the CoREMOF database and predicted several promising MOF for CO$_2$ capture. Moreover, we carry out GCMC calculations for a few screened candidates and compare them with DL predictions showing promising trends. This study demonstrates some of the key strength for using GNN over usual conventional methods. We also discuss some of the weaknesses of the model that motivate further development for the GNN model. Our work also establishes that GNN models can not only be successfully applied for molecules and solids but MOF structures as well.

\section*{acknowledgements}
K.C. thanks National Institute of Standards and Technology for computational and funding support.

\section*{conflict of interest}
The authors declare no conflict of interest.

\section*{Supporting Information}

Supporting Information is available from the Wiley Online Library or from the author. A list of DL based screened MOF candidates in provided in the supporting information.

\printendnotes

\bibliography{sample}



\end{document}